\documentclass{article}

\usepackage{arxiv}

\usepackage[utf8]{inputenc} % allow utf-8 input
\usepackage[T1]{fontenc}    % use 8-bit T1 fonts
\usepackage{hyperref}       % hyperlinks
\usepackage{url}            % simple URL typesetting
\usepackage{booktabs}       % professional-quality tables
\usepackage{amsfonts}       % blackboard math symbols
\usepackage{nicefrac}       % compact symbols for 1/2, etc.
\usepackage{microtype}      % microtypography
\usepackage{lipsum}

\usepackage{graphicx} 
\usepackage{svg}
\usepackage{float}
\usepackage{hyperref}
\usepackage{mathptmx}

\title{Salt parameterization can drastically affect the results from classical atomistic simulations of water desalination by MoS$_2$ nanopores}

\author{
 Jo\~ao P. K. Abal\\
  Institute of Physics\\
  Federal University of Rio Grande do Sul\\
  Brazil, Porto Alegre, 91501-970\\
  \texttt{joao.abal@ufrgs.br} \\
   \And
  José Rafael Bordin \\
  Department of Physics\\
  Institute of Physics and Mathematics\\
  Federal University of Pelotas\\
  Brazil, Pelotas, Rua dos Ip\^es,96050-500\\
  \texttt{jrbordin@ufpel.edu.br} \\
  \And
  Marcia C. Barbosa\\
    Institute of Physics\\
  Federal University of Rio Grande do Sul\\
  Brazil, Porto Alegre, 91501-970\\
  \texttt{marcia.barbosa@ufrgs.br} \\
}

\begin{document}
\maketitle

\begin{abstract}
Water scarcity is a reality in our world, and scenarios predicted by leading scientists in this area indicate that it will worsen in the next decades. However, new technologies based in low-cost seawater desalination can prevent the worst scenarios, providing fresh water for humanity. With this goal, membranes based in nanoporous materials have been suggested in recent years. One of the materials suggested is MoS$_2$, and classical Molecular Dynamics (MD) simulation is one of the most powerful tools to explore these nanomaterials. However, distinct Force Fields employed in  MD simulations are parameterized based on distinct experimental quantities. In this paper, we compare two models of salt that were build based on distinct properties of water-salt mixtures. One model fits the hydration free energy and lattice properties, the second fits the crystal density and the density and the dielectric constant of water and salt mixtures. To compare the models, MD simulations for salty water flow through two nanopores sizes were used -- one pore big enough to accommodate hydrated ions, and one smaller in which the ion has to dehydrate to enter, and two rigid water models from the TIP4P family -- the TIP4P/2005 and TIP4P/$\epsilon$.  Our results indicate that the water permeability and salt rejection by the membrane are more influenced by the salt model than by the water model, especially for the narrow pore. In fact, completely distinct mechanisms were observed, and they are related to the characteristics employed in the ion model parameterization. The results show that not only the water model can influence the outcomes, but the ion model plays a crucial role.
\end{abstract}

% keywords can be removed
\keywords{Desalination \and Nanoporous Membrane \and Nanofluidics \and Water Models \and Salt Models \and Molecular Dynamics}

\section{Introduction}

%--- The water problem and the desalination

One of the greatest challenges of our time is concerned with water scarcity. Currently, our freshwater resources are dwindling at an unprecedented rate due to a high imbalance between clean water demand and total supply~\cite{wwreport2019}.
In the face of growing water scarcity, it is critical to understand the potential of salty water desalination as a long-term water supply option~\cite{JONES20191343}. The Reverse Osmosis (RO) system is considered the leading desalination process and the best available option in terms of energy consumption~\cite{QASIM201959}. This technique is based on a membrane separation method. However, the energy and monetary cost of RO with the current membranes are high mainly because of the membrane fouling phenomena. The new and promising technology is to use membranes made of nanomaterials~\cite{TEOW20192} as graphene~\cite{doi:10.1021/nl3012853,COHENTANUGI201559}, and molybdenum disulfide~\cite{doi:10.1038/ncomms9616}, which shows improved permeability potential at exceptional separation capability.

 The key component of a good membrane is the balance between water permeability and salt rejection, in such a way that the next-generation membranes need to be very selective~\cite{boretti}. Molecular dynamics simulations are a powerful tool to mimic a reverse osmosis system at nanoscale~\cite{zhu-2014}. It helps us to get insights in design new membranes materials and better understand the water-salt-nanopore relationship~\cite{COHENTANUGI201559}. The water flux throughout the membrane can be generally related to its specific permeability by the following expression: A$_m$ = $\phi$/(P - $\Pi$), in which A$_m$ is the membrane specific permeability, $\phi$ is the water flux, $P$ is the applied pressure and $\Pi$ is the osmotic pressure. All these quantities can be obtained or controlled by designing the system for molecular dynamics simulations.

%--- Recent membranes advances

Graphene based nanomembranes are well known in the literature~\cite{doi:10.1021/nl3012853, AGHIGH2015389} and have been extensively studied, showing its efficiency in water desalinations~\cite{TEOW20192, surwade15}.
Another promising material is nanoporous molybdenum disulfide (MoS$_2$). Their efficiency has been investigated by molecular dynamics simulations~\cite{li-acsnano2016,kohler-jcp2018,perez-apl2019,kou-pccp2016} and experimental works~\cite{li-nl2019,wang-nl2017,zhou-nl2013,doi:10.1021/acsnano.7b05124}, showing that the combination of hydrophobic and hydrophilic sites in the nanopore can increase the desalination performance.

Molecular dynamics simulations are a suited theoretical approach to understand the physics behind nanofluidic systems once it allows for probing the microscopic behavior of atoms while performing timescale feasible simulations~\cite{gravelle:hal-02375018}. In addition, to represent the system computationally one has to face the challenge of design a model capable to encode the main physics of the problem. Said that the model chosen to represent the interactions of the atoms is the seed in which the whole dynamics arise following the classical equations of motion. In the specific case of classical atomistic Molecular Dynamics simulations, most of the Force Fields use simple additive, nonpolarizable, and pairwise potential for atomic interaction~\cite{Cordo09, doi:10.1021/ja00316a012,doi:10.1063/1.1570405,Underwood2018,doi:10.1063/1.1387447,doi:10.1063/1.4728163}. In the case of water, rigid nonpolarizable models are extensively employed in simulations of bulk~\cite{doi:10.1080/08927022.2018.1511903} and nanoconfined~\cite{Losey19,KOHLER201954,doi:10.1021/acs.jpcc.8b00112,C7CP02058A,KOHLER2018331} systems. Efforts has been done to include polarization in classical simulations~\cite{doi:10.1021/ct900576a, doi:10.1146/annurev-biophys-070317-033349, doi:10.1021/ja00007a021, BordinPol}, but nonpolarizable salt and water remain as the main models in MD simulations of desalination. 

Another issue that has to be handle with care relies in the optimization of specific ion parameters 
for specific water models. As D\"opke and co-authors have recently showed~\cite{doi:10.1063/1.5124448}, salt models optimized for SPC/E and TIP3P water
can lead to wrong predictions when dissolved in TIP4P/2005 water. This is relevant once the TIP4P/2005~\cite{doi:10.1063/1.2121687} model is one of the best and most employed rigid water models. 

In recent works about water desalination by nanopores~\cite{doi:10.1038/ncomms9616, doi:10.1063/1.4866643, doi:10.1021/acs.nanolett.5b04089, perez-apl2019, C9CP04364K, doi:10.1063/1.4975690, doi:10.1021/jp512358s} the ion model proposed by Joung and Cheatham~\cite{doi:10.1021/jp8001614} has been employed. This model,  which will be referred to as NaCl/J, was parameterized based on the hydration free energies of the solvated ions and lattice parameters of salt crystals and has a good agreement with several experimental studies. 
These parameters were optimized in combination with some of the most classical water models, as SPC/E,  TIP3P,  or  TIP4P/Ew  water. Also, as Liu and Patey~\cite{doi:10.1063/1.4975690} and  D\"opke and ~\cite{doi:10.1063/1.5124448} discuss in their works, the ion parameters optimized for TIP4P/Ew can be transferred to TIP4P/2005 water without lost of accuracy.  
On the other hand, the dielectric discontinuity of water near interfaces and nanopores plays a crucial role in salt behavior~\cite{doi:10.1021/jp011058i, Levin_2002, PhysRevE.85.031914}. Recently, Fuentes and Barbosa proposed the NaCl/$\epsilon$ model~\cite{doi:10.1021/acs.jpcb.5b12584}. This model was parameterized to reproduce the experimental values of density of the crystal and the density and dielectric constant of the mixture of the salt with water at a diluted solution when combined with the TIP4P/$\epsilon$ rigid water model. To reproduce these properties and correct for the nonpolarizability of the model they propose a screening factor in the Coulomb interaction - usually, nonpolarizable models are parametrized based in the Lennard-Jones (LJ) potential parameters only.

In this paper, we answer the question about how distinct ionic models influence the MoS$_2$ membrane water desalination study. To do so we compare a model of ions constructed based on hydration and crystal properties, and a model constructed to reproduce the density and dielectric constant of water and salt mixtures. Our paper is organized as follows. In Section 2 we introduce our model and the details about the simulation method. In Section 3 we show and discuss our results, and the conclusions are presented in Section 4.

\section{Methods and Simulational Details}

\begin{figure}[h!]
\centering
\includegraphics[width=8cm]{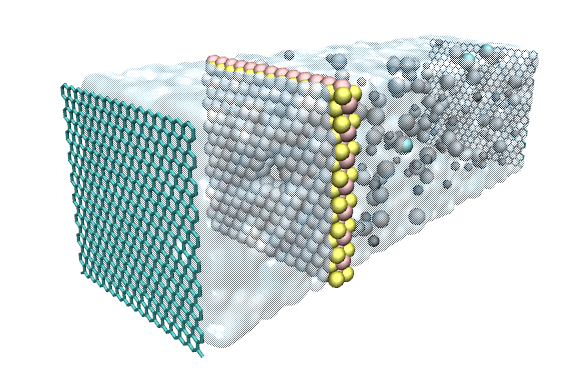}
\caption{The illustration of a typical desalination box. Image created using the VMD software~\cite{HUMP96}.}
\label{fig1}
\end{figure}

One of the most employed methodology to simulate the saltwater desalination process in MD simulations~\cite{li-acsnano2016,kohler-jcp2018,perez-apl2019,kou-pccp2016} is based on the creation of a box with the membrane located between two confined reservoirs, one of pure water and another one with saltwater, as we show in Figure~\ref{fig1}. The reservoirs can be confined by graphene barriers, for example. This barriers are used as pistons to control the confined solution pressure.  To mimic the water driven force throughout the membrane, one has to apply different pressures in each reservoir.

Molecular dynamics were performed using the LAMMPS package~\cite{PLIMPTON19951}. The initial system in between the graphene barriers has 4 x 4 x 125 nm in x,y, and z, respectively. The periodic boundary conditions were used in all directions. By doing that, one has to use a simulation box large enough in z direction in order to guarantee the molecules don't interact with each other across that boundary. The saltwater used has almost 1 mol/L of solute concentration (170 ions for 4930 water molecules), higher than the average seawater salinity of 0.6 mol/L. The pure water side contains 1550 molecules. 

The salt and water Lennard-Jones parameters and charges were taken from the papers that proposed each model: the NaCl/$\epsilon$ model~\cite{doi:10.1021/acs.jpcb.5b12584}, the NaCl/J~\cite{doi:10.1021/jp8001614} model, the Tip4p/$\epsilon$~\cite{FUENTESAZCATL201686} model and the Tip4p/2005~\cite{doi:10.1063/1.2121687}. The parametrization of a reactive many-body potential was used as LJ parameters and charges values for molybdenum and sulfur, as proposed by Kadantsev and Hawrylak ~\cite{KADANTSEV2012909}. The carbon parameters from the graphene piston was taken from the seminal work on confined water by Hummer and co-workers~\cite{Hummer2001}. For simplicity, the MoS$_2$ membrane remained fixed in all simulation, and the graphene sheet has freedom only in the flow direction.
For the non-bonded interactions, the Lorentz-Berthelot mixing rules were employed. The long-range electrostatic interactions were calculated by the particle-particle-particle mesh method and the LJ
cutoff distance was 1 nm. The SHAKE algorithm was used to maintain the water molecules rigid.

\begin{figure}[h!]
\centering
\includegraphics[width=8cm]{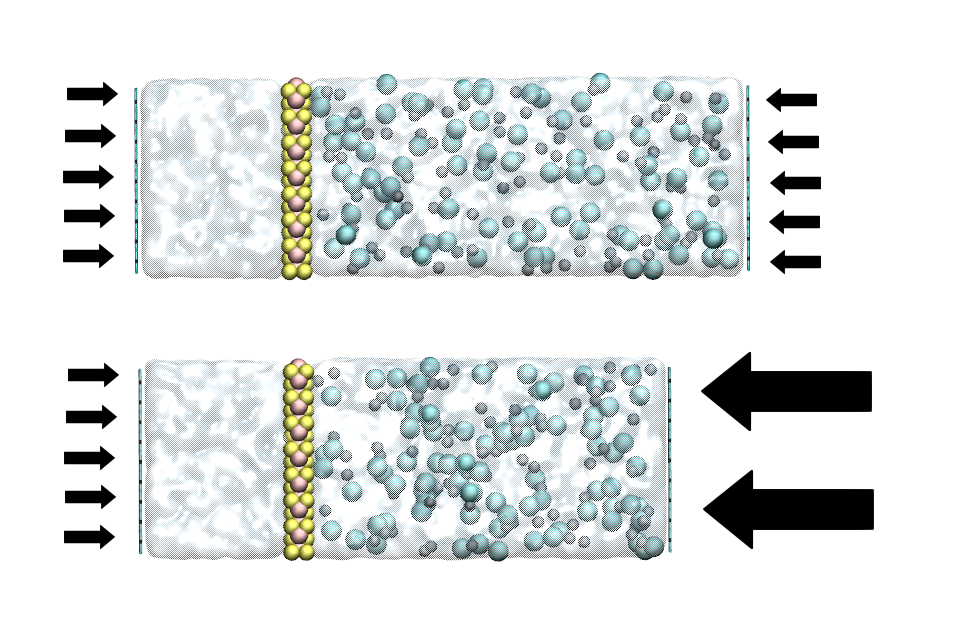}
\caption{Schematic depiction of the non-equilibrium scheme: the pressure on the left graphene sheet is constant and equal to the atmospheric pressure, 1 bar, while distinct higher values are applied in the right graphene sheet, ranging from 1000 to 10000 bar.}
\label{fig2}
\end{figure}

First, each energy simulation was minimized for 0.5 ns on NVE ensemble. It means that the graphene sheets are freeze at that time. After that, the simulations were equilibrated with a constant number of particles, pressure, and temperature (NPT) ensemble for 1 ns at 1 bar and 300 K, as illustrated in the Figure~\ref{fig2}-up. The pressure control was made by leaving the graphene pistons free to move in the z-direction and applying a force in each carbon atom in order to produce the desired ambient pressure. After some simulation steps, the solution equilibrates and the piston pressure reaches the equilibrium density at ~1 g/cm$^3$. Then, the graphene sheets were frozen and 2 ns simulations in NVT ensemble were performed to further equilibrate the system. The Nosé-Hoover thermostat was used with a time constant of
0.1 ps~\cite{doi:10.1063/1.447334,PhysRevA.31.1695}.

Next, the nanopore was opened by removing the desired atoms of molybdenum and sulfur in order to maintain the membrane charged neutral. The two nanopores studied has 0.74 nm and 0.97 nm of diameter respectively. The nanopores sizes were calculated by using the center-to-center distance of atoms. Finally, the external pressure was applied on the feed side and the non-equilibrium running was carried out for 10 ns, as illustrated in the Figure~\ref{fig2}-bottom. Each run was averaged over 3 sets of simulations with different initial thermal velocity distributions. The feed pressures range from 1000,2500,5000 to 10000 bars. We used such high pressures for statistical purposes. 

\section{Results and Discussions}

Distinct models can lead to water flow rates in nanopores because the different number of sites, flexibility, partial charges, and LJ parameters can strongly change the observed flow~\cite{Losey19}. In a similar way, the ion parameters can affect the ionic blockage and binding in biological~\cite{Ash06, Cordo09} and synthetic nanopores~\cite{Beck04, He13, Hsu17, Abraham17}. In fact, a considerable amount of factors affects the ion entry in nanopores~\cite{Beck04, Beck01}. In order to investigate the role of the screening, we evaluate the water and ion flow trough nanopores with diameters of 0.97 nm or 0.74 nm using two distinct water and ion model. For the wider diameter, the ion enters in the nanopore screened by water, while for the smaller diameter the ion has to strip out the water in order to penetrate the pore. These two cases allow us to compare not only the model effect but the screening effect.

\begin{figure}[H]
\centering
\includegraphics[width=8cm]{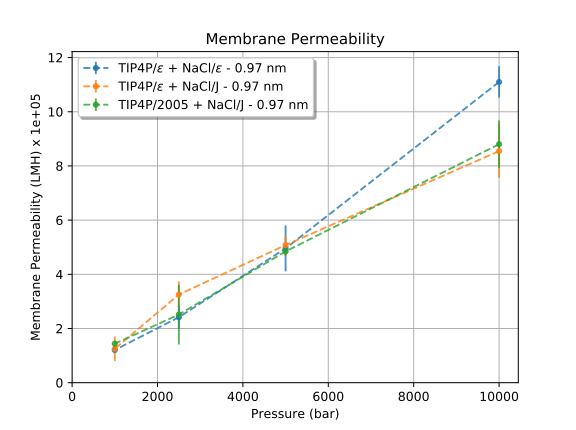}

\caption{Membrane water permeability for distinct combinations of water and salt models and nanopore with 0.97 nm diameter and. Error bars are the deviation from the mean value - errors bars smaller than the point are not shown.}
\label{fig3a}
\end{figure}

\begin{figure}[H]
\centering
\includegraphics[width=8cm]{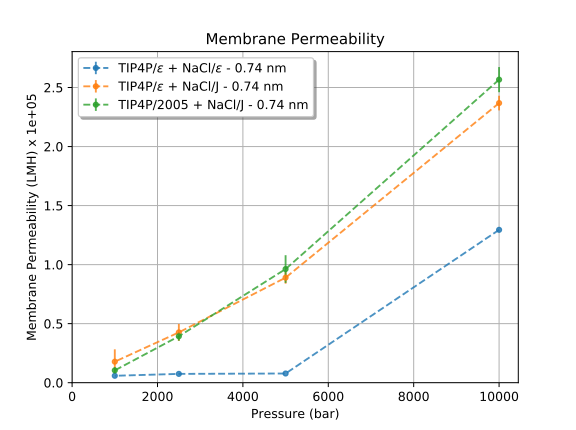}
\caption{Membrane water permeability for distinct combinations of water and salt models and nanopore with 0.74 nm diameter. Error bars are the deviation from the mean value - errors bars smaller than the point are not shown.}
\label{fig3b}
\end{figure}

Despite the fact that TIP4P/2005 and TIP4P/$\epsilon$ have both 4 charged sites, we can expect that the distinct values in its parameters can affect the permabillity of pure water trough nanopores. However, as Losey and co-workers have shown in a recent work~\cite{Losey19}, TIP4P and TIP4P/2005 water models have the similar flow rates. In agreement with this result, our simulations shows that when the same model of salt is employed, the membrane permeability for both TIP4P/2005 and TIP4P/$\epsilon$ is approximately the same -- the differences are smaller than the error bar, as we can see in the Figure~\ref{fig3a,fig3b}, from both nanopore sizes. Changing the salt model to NaCl/$\epsilon$ affects the water permeation in the widest nanopore at the higher values of pressure. As we can see in Figure~\ref{fig3a}, when the applied pressure is 10000 bar the combination of TIP4P/$\epsilon$+NaCl/$\epsilon$ shows a higher water flow rate. On the other hand, the water permeability is small for this combination in the case of nanopores with 0.74 nm diameter, as shown in Figure~\ref{fig3b}. Actually, the permeability is small and approximately constant for the three smallest values of applied pressure - it is necessary a huge pressure gradient to create a bigger water flow through the nanopore.

\begin{figure}[H]
\centering
\includegraphics[width=8cm]{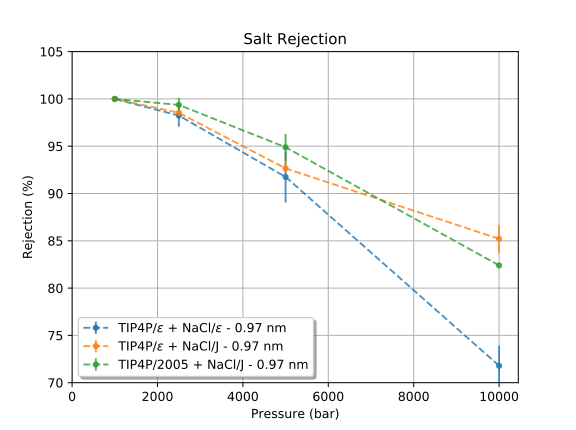}
\caption{Salt rejection for distinct combinations of water and salt models and nanopore with 0.97 nm diameter. Error bars are the deviation from the mean value - errors bars smaller than the point are not shown}
\label{fig4a}
\end{figure}

\begin{figure}[H]
\centering
\includegraphics[width=8cm]{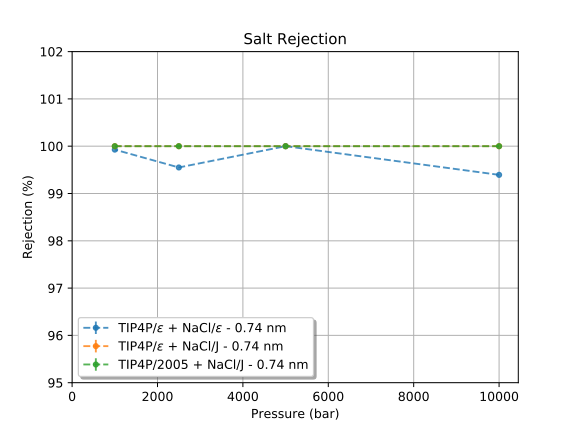}
\caption{Salt rejection for distinct combinations of water and salt models and nanopore with 0.74 nm diameter. Error bars are the deviation from the mean value - errors bars smaller than the point are not shown}
\label{fig4b}
\end{figure}

The distinct values of water permeation for each combination of water and salt model, as well for each nanopore size, Figure~\ref{fig3a,fig3b} are related to distinct salt rejection. As we show in the Figure~\ref{fig4a}, the salt rejection in the wider pore decreases with the applied pressure, and the NaCl/$\epsilon$ has the smallest rejection at the higher pressure - in agreement with the higher water permeability. For the narrow pore, the system with NaCl/J salt shows 100$\%$ of rejection as shown in the Figure~\ref{fig4b} and in agreement with our previous results~\cite{kohler-jcp2018}. In the case of the Na/$\epsilon$ salt model, a few ions can cross the pore. The membrane specific permeability obtained for each mixture and nanopore sizes are summarized in Table~\ref{table1}.

\begin{table}\vspace{1cm}
\centering
\begin{tabular}{c c c}
\hline
\textbf{Mixture} & A$_m$ [LMH/bar] & Diameter [nm] \\
\hline
    TIP4P/$\epsilon$ - NaCl/$\epsilon$ & 108.2 (17.5) & 0.97 \\
    TIP4P/$\epsilon$ - NaCl/J & 104.1 (28.6) &0.97   \\
    TIP4P/2005 - NaCl/J & 118.7 (25.8) & 0.97\\
    TIP4P/$\epsilon$ - NaCl/$\epsilon$ & 5.9 (5.1) &  0.74 \\
    TIP4P/$\epsilon$ - NaCl/J & 17.2 (5.7) &  0.74 \\
    TIP4P/2005 - NaCl/J & 18.4 (5.8) & 0.74  \\
\hline
\end{tabular}
\caption{The membrane specific permeabilities (A$_m$) obtained for such nanopore sizes considering the  nanopore density of 6.25 cm$^{-2}$.}
\label{table1}
\end{table}\vspace{1cm}

\begin{figure}[H]
\centering
\includegraphics[width=8cm]{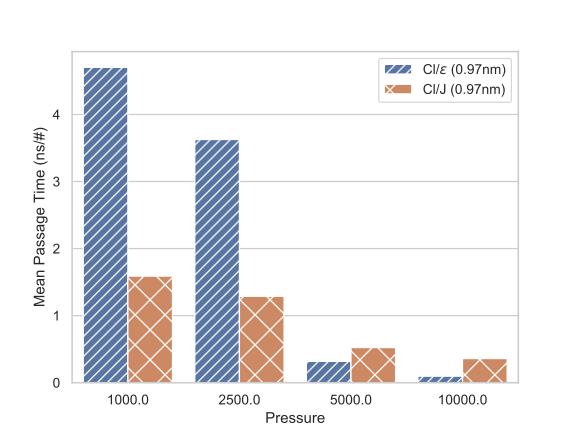}
\caption{Mean passage time (MPT) versus applied pressure for different models and nanopore diameters.  Cl/$\epsilon$ and Cl/J MPT in 0.97 nm diameter are compared. Although the anions remains a considerable amount of time inside the nanopore, the water still can  flow as shown in Figure~\ref{fig8}.}
\label{fig5a}
\end{figure}

\begin{figure}[H]
\centering
\includegraphics[width=8cm]{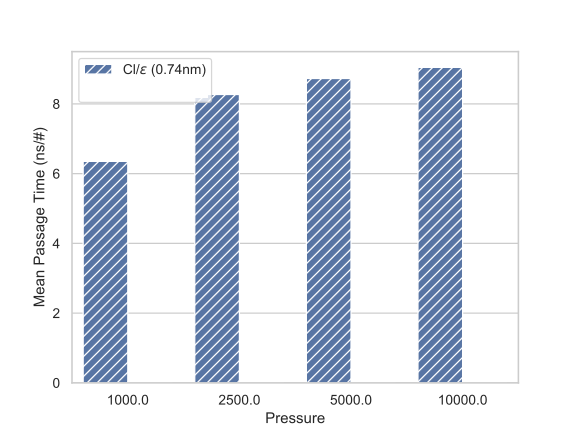}

\caption{Mean passage time (MPT) versus applied pressure for different models and nanopore diameters. Cl/$\epsilon$ MPT in 0.74 nm diameter are shown. The Cl/$\epsilon$ remains almost the total simulation time blocking the nanopore. In contrast, the Cl/J don't enter in the nanopore therefore its not shown.}
\label{fig5b}
\end{figure}

To understand the water and ions permeation trough the pore, we evaluate the Mean Passage Time (MPT) of the different ion models through the nanopore with the two studied diameters. As we show in Figure~\ref{fig5a,fig5b}, the Cl/$\epsilon$ anions are responsible for the nanopore blockage when this model is employed. Despite the case of 10000 bar of applied pressure, in all other cases, the chlorine takes a long time to pass the pore and therefore is the ion blocking the pore. Even for the wider nanopore, the blockage time is relevant at lower pressures, with the Cl anion remaining almost 5 ns, or half of the production time, inside the pore. On the other hand, the Cl/J anion remains short times inside the nanopore with 0.97 nm diameter, which explains the higher water permeability and smaller ionic rejection and never enters the smallest pore, as shown in Figure~\ref{fig5b}.

\begin{figure}[H]
\centering

\includegraphics[width=8cm]{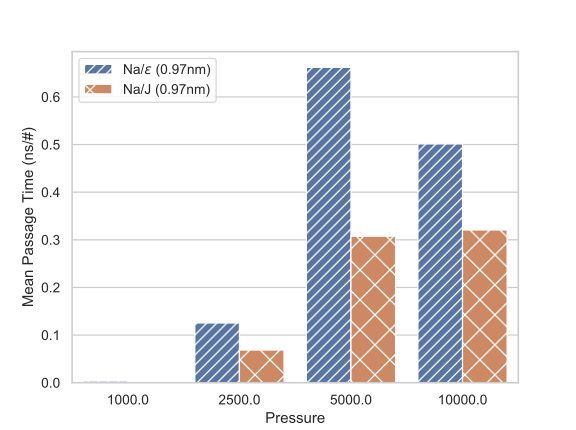}
\caption{Mean passage time (MPT) versus applied pressure for different models and nanopore diameters. Na/$\epsilon$ and Na/J MPT in 0.97 nm diameter are compared. The Na/$\epsilon$ and Na/J case for 0.74 nm diameter are not shown because they never enter in that nanopore.}
\label{fig5c}
\end{figure}

For the smaller pore, distinct mechanisms are responsible for the ionic rejection is interesting. For the NaCl/$\epsilon$ the pore is blocked by the chlorine anion (see Figure~\ref{fig5b} and Figure~\ref{fig8}), while for the NaCl/J model the chlorine never enters the pore. In addition, the sodium cations takes a short time to pass the wider nanopore, as illustrated in Figure~\ref{fig5c}.
At this point, is relevant to emphasize that the ionic passage trough small pores has two main events~\cite{PhysRevE.85.031914}: first the ion must hit the pore, secondly it need to have enough energy to overcome the energetic penalty related to leave the bulk, enter the pore with a distinct dielectric constant, and cross the pore to the bulk again. The first process is a classical problem from statistical mechanics, depending mainly in the system density and pore area~\cite{LEVIN2004543, PhysRevLett.120.120601}. In the second process the penalties can depend on the nanopore size, ion hydration, ion charge, pore chemical characteristics and pore geometry~\cite{Beck04, PhysRevE.85.031914}. Therefore, we will now analyze how the salt models properties influences the ion translocation event.


The translocation process of the ions is analyzed by the Potential of Mean Force (PMF) which quantifies the passage of the ion from bulk, enters the pore, crosses it and leaves to the other bulk region.
Here, the PMF calculations were obtained by preparing a set of different systems in which one specific ion was freeze in a position along the $z$ direction aligned with the center of the nanopore. At this specific position, we run 0.5 ns of simulation, the time required for the salt and water around the ion achieve the equilibrium, with the pore closed and without pressure gradient. Then, the external pressure is increased to 1000 bar and the nanopore is opened. With the ion still fixed in space, we evaluated the force felt by this ion for another 0.5 ns. After that, we increased the $z$ position of the ion by a $\delta z = 0.5$ \AA, repeat the steps in the equilibrium and in the non-equilibrium, and so on until ion crosses the pore to the other bulk region. After that, the PMF was obtained by the integration of the total mean force along the $z$ direction. The PMF calculations were done using only the TIP4P/$\epsilon$ water for two complementary reasons. First, the permeation seems to be more sensitive to the ion model rather than to the water model. Second, the electrostatic barrier related to the dielectric discontinuity from the bulk water to the nanopore region is relevant, and this water model was parameterized to provide the correct value of bulk water dielectric constant~\cite{FUENTESAZCATL201686}. In the same spirit, the NaCl/$\epsilon$ was parameterized to reproduce the dielectric constant of the mixture of the salt with water at a diluted solution~\cite{doi:10.1021/acs.jpcb.5b12584}.

\begin{figure}[H]
\centering
\includegraphics[width=8cm]{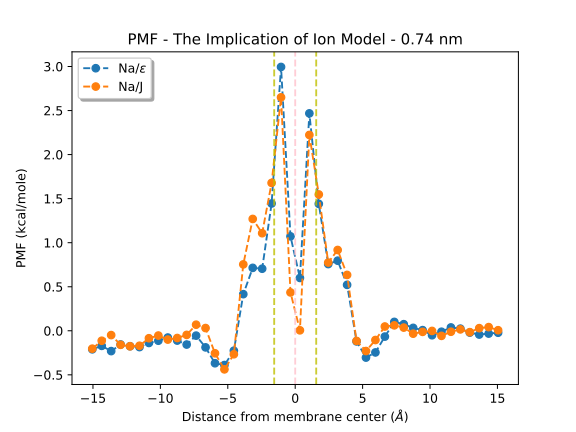}
\caption{PMF inside nanopores with 0.74 nm diameter for sodium ions. The central vertical pink dashed line represents the molybdenum layer position, and the vertical yellow dashed lines around it the sulfur layers position.}
\label{fig6a}
\end{figure}

\begin{figure}[H]
\centering
\includegraphics[width=8cm]{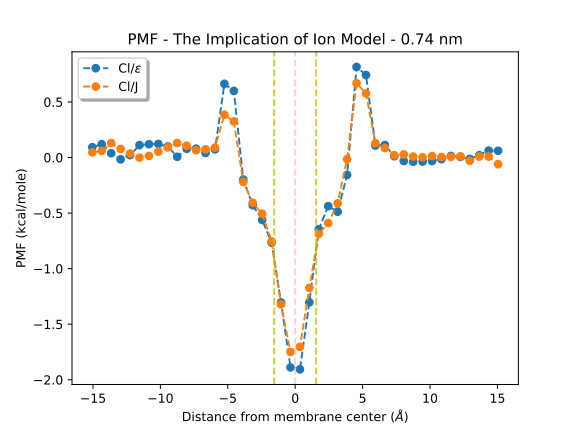}
\caption{PMF inside nanopores with 0.74 nm diameter for chlorine ions. The central vertical pink dashed line represents the molybdenum layer position, and the vertical yellow dashed lines around it the sulfur layers position.}
\label{fig6b}
\end{figure}

As we show in the Figure~\ref{fig6a}, the energetic penalty for a sodium ion to leave the bulk and to enter the nanopore with diameter 0.74 nm is more than 5 times the thermal energy at 300 K, $k_B T \approx 0.6$ kcal/mole. This explains why we have to apply a huge pressure to observe a Na cation inside this small pore. On the other hand, the energy barrier for a Cl anion is much smaller, comparable with the thermal energy, for both models. Therefore the anion can penetrate the pore only due to thermal fluctuations at room temperature. However, the central well has a deepness of 4 to 5 times $k_B T$, created by the attraction with the central layer of positively charged molybdenum. Then the Cl$^-$ gets trapped. This, however,  does not explain why the Cl/$\epsilon$ enters and block the nanopore, while the Cl/J never leaves the bulk to the pore.

\begin{figure}[H]
\centering
\includegraphics[width=8cm]{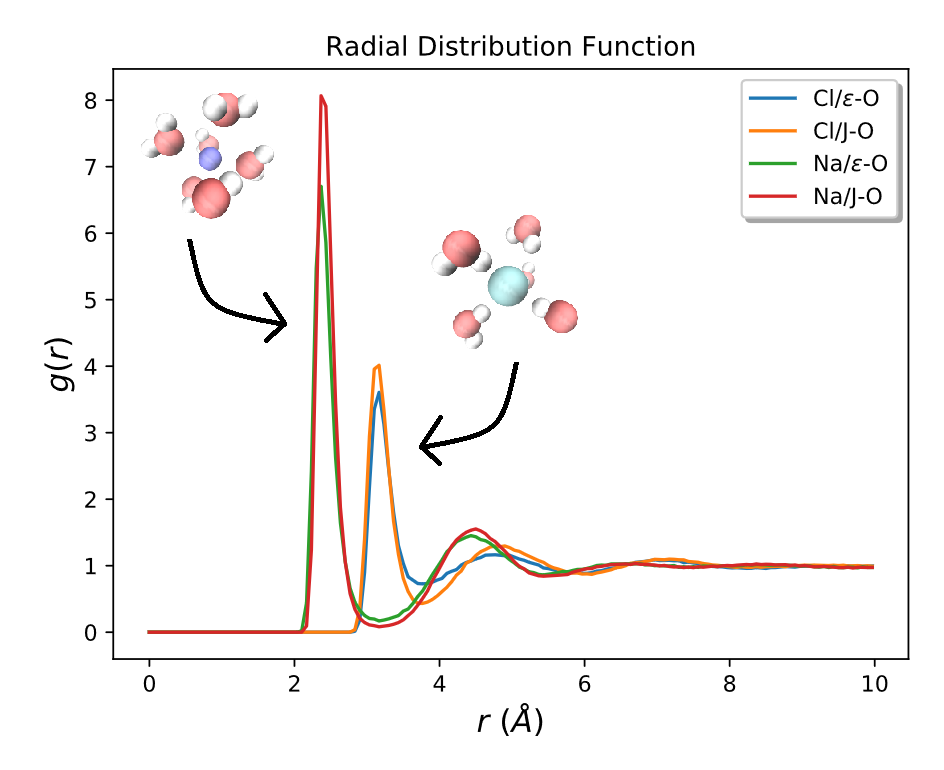}
\caption{Bulk ion-oxygen radial Distribution at 300k and 1000 bar.}
\label{fig7a}
\end{figure}

\begin{figure}[H]
\centering
\includegraphics[width=8cm]{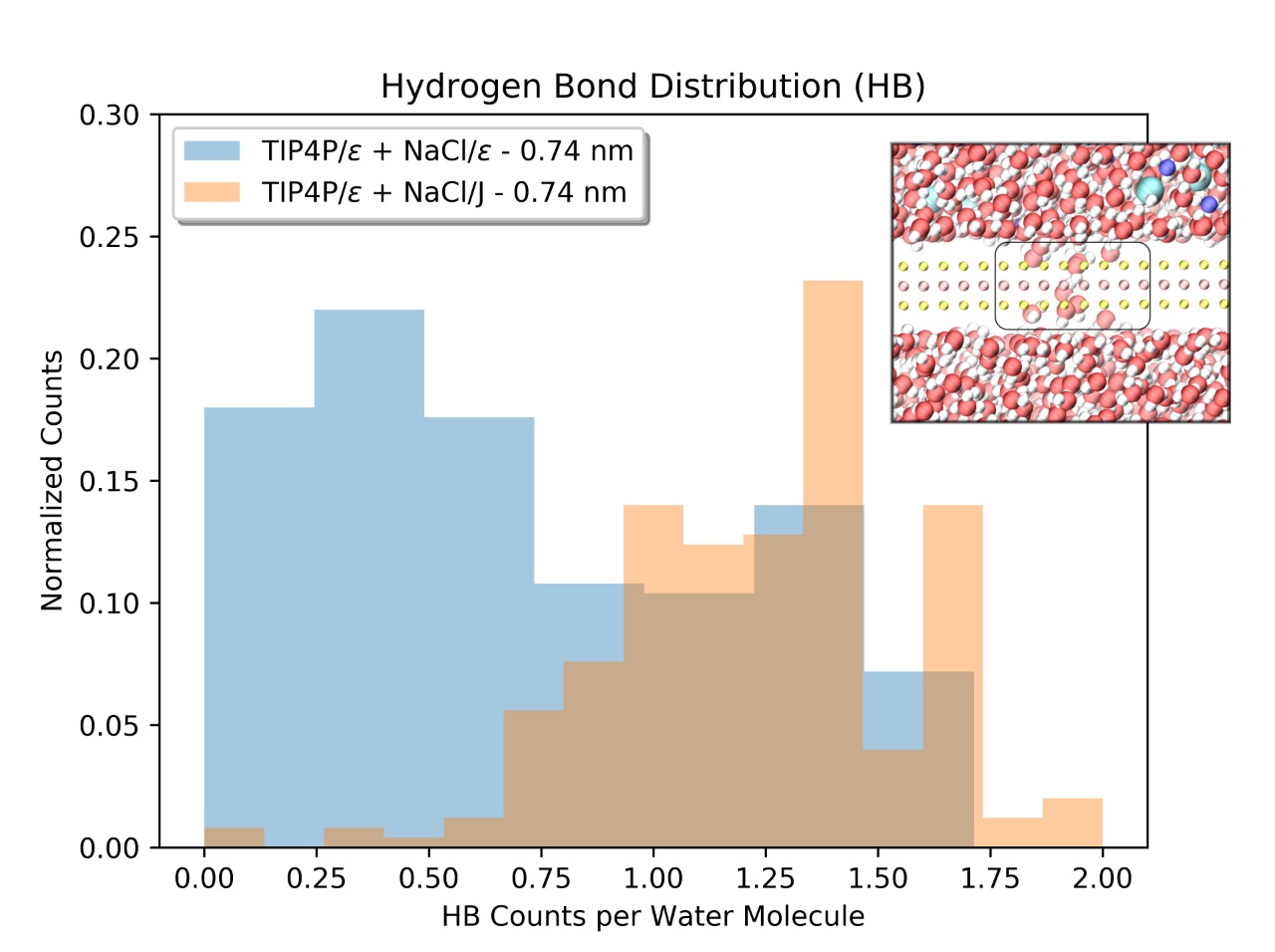}
\caption{Hydrogen bond (HB) distribution near the pore. In the inset we show the region considered to evaluate the distribution.}
\label{fig7b}
\end{figure}

The reason for distinct blockage when the two models are compared for the nanopore with diameter 0.74 nm is that the NaCl/J model is more hydrated than the NaCl/$\epsilon$ model. The different screening factors employed affect water distribution around the ions. In Figure~\ref{fig7a} we show the radial distribution function, $g(r)$ of the oxygen atoms of the water molecules around the distinct species of ions. For the sodium ions, the peaks are smaller for the Na/$\epsilon$ than for the Na/J model. For the Na ion, however, the peaks distances are the same and the water structure around the Na is independent of the water model. On the other hand, for the chlorine ions, not only the peaks for the Cl/$\epsilon$ are smaller, but the water seems more disordered: the depletion between the first and second hydration layer is shallower, and the curve is almost flat after this second peak.  This is confirmed when we evaluate the hydrogen bond (HB) distribution near the pore,
shown in Figure~\ref{fig7b} (the HB distribution was obtained by following distance and angular criteria considering the $r_{O-O} < 3.5$~\AA~and $\theta_{OH-O} < 30^{\circ}$)~\cite{C7SC04205A}. As we can see, for the NaCl/$\epsilon$ model more then 60\% of the water molecules form less than one HB in average. On the other hand, when the NaCl/J model is employed each water molecule forms more than one hydrogen bond. Therefore the salt model affects not only the ion wettability, but can effectively change the water HB network. Then, due the higher hydration and the higher number of HB by water molecule, the the Cl/$\epsilon$ can strip out this water easily in comparison to the Cl/J model and enter the channel. This "water striping" is essential, since the small nanopore diameter of 0.74 nm makes impossible to hydrated ion penetrate -- as we have observed and show in the upper snapshot~\ref{fig8}.

\begin{figure}[H]
\centering
\includegraphics[width=8cm]{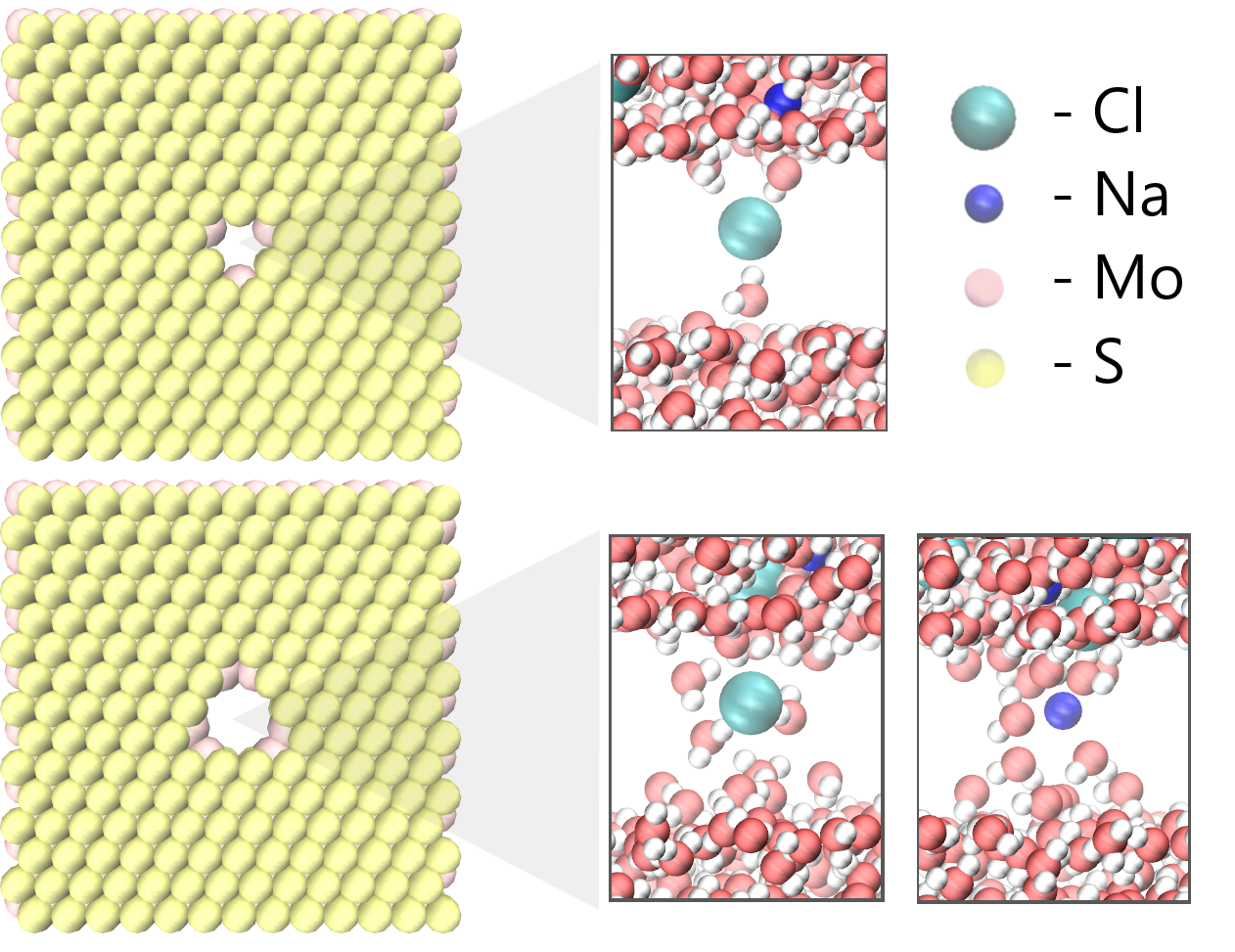}
\caption{Snapshots of the simulation showing that for the smaller nanopore (upper snapshot) only dehydrated Cl/$\epsilon$ ions can penetrate the pore, while for the nanopores with diameter 0.97 nm we observe permeation of both hydrated ionic species and models.} 
\label{fig8}
\end{figure}

\begin{figure}[H]
\centering
\includegraphics[width=8cm]{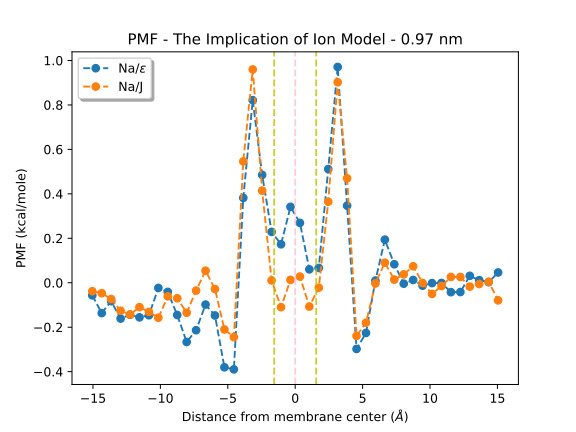}
\caption{PMF inside nanopores with 00.97 nm diameter for sodium ions. The central vertical pink dashed line represents the molybdenum layer position, and the vertical yellow dashed lines around it the sulfur layers position}
\label{fig9a}
\end{figure}

The nanopore with diameter 0.97 nm is wide enough to accommodate hydrated ions, as we show in the lower panel of Figure~\ref{fig8}. This hydration makes the dielectric discontinuity between the bulk and the pore small, decreasing the energetic penalty for the ion current through the nanopore. The barrier, illustrated in the Figure~\ref{fig9a}, for the sodium ions is now smaller than twice the thermal energy. Therefore, the ions can cross the pore as the pressure increase, as we have observed in Figure~\ref{fig4a}. Also, the depth of the well for the Na/$\epsilon$ is small -- so we observe a smaller ion rejection. 
This is also a consequence of the screening in the Coulomb interaction between the salt and the pore ions, which should rule the PMF once the ions are hydrated and the dielectric discontinuity is small.
And, for this nanopore, we do not observe a significant difference in the PMF for both Cl models, what can explain why the mean passage times of the chlorine ions for both models are comparable in the wide pores, especially at high pressure, as shown in the Figure~\ref{fig5a}.
In a similar way, the water HB distribution near the pore region is similar for both ion model, as the Figure~\ref{fig9c} show. The small difference, whit the  NaCl/$\epsilon$ leading to less HBs, can also be associated with the small ion rejection observed for this salt model.

\begin{figure}[H]
\centering
\includegraphics[width=8cm]{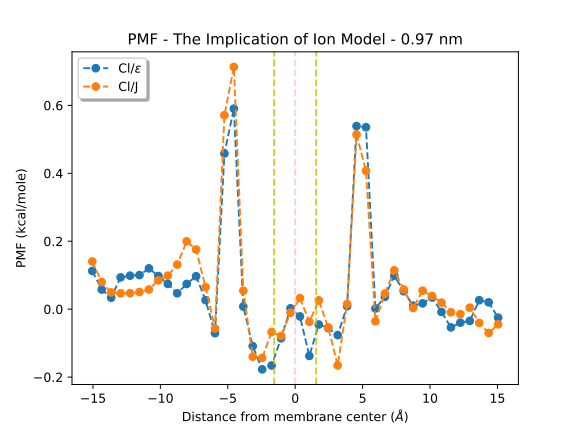}
\caption{PMF inside nanopores with 00.97 nm diameter for chlorine ions. The central vertical pink dashed line represents the molybdenum layer position, and the vertical yellow dashed lines around it the sulfur layers position.}
\label{fig9b}
\end{figure}

\begin{figure}[H]
\centering
\includegraphics[width=8cm]{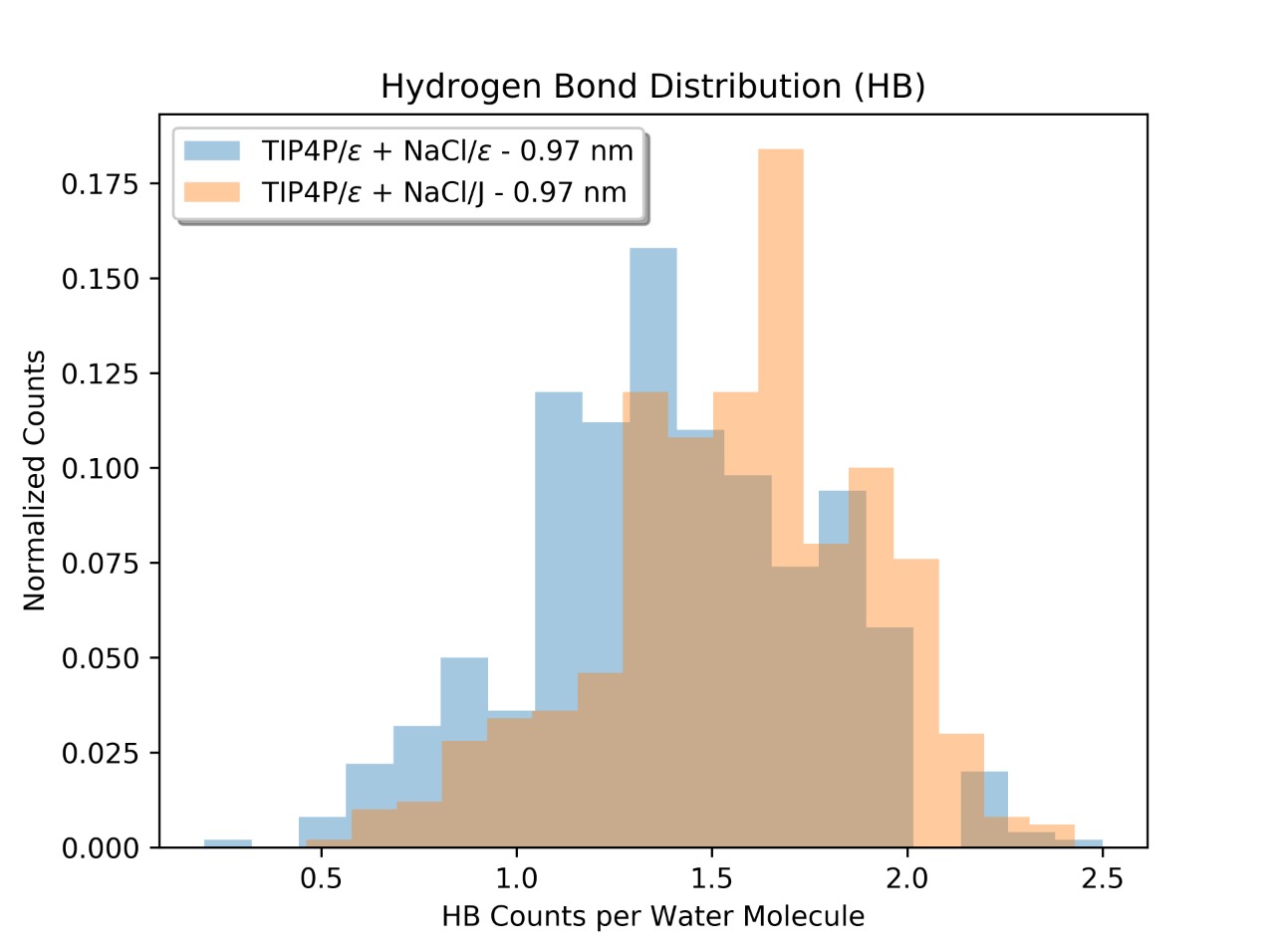}
\caption{Hydrogen bond distribution near the pore.}
\label{fig9c}
\end{figure}

These results indicate that the effect of ion rejection depends on the ion model applied. In one case, using the model that did not reproduce properly the dielectric constant of water and salt mixtures, the rejection is due to the dielectric discontinuity and the energetic penalty associated with the ion dehydration. In the other case, employing the model that reproduces the bulk dielectric constant of salt and water mixtures, the pore is blocked by the chlorine ion. Obviously, the second case is not interesting since it does not allow the water permeation through the pore. This blockade was observed in experiments for single-layer graphene membranes~\cite{surwade15} also been suggested by DFT modeling of functionalized graphene nanopores~\cite{Guo14}. This effect is well known for polymeric membranes~\cite{doi:10.1081/SS-120039343} and it is a big challenge on reverse osmosis engineering.
However, it was not reported experimentally or by simulations for MoS$_2$ membranes so far we know. These results indicate that extensive research has to be done, especially experimental studies, to see if there is ionic blockage or not for MoS$_2$ small nanopores once membrane fouling control is one of the most important performance parameters for next-generation membrane materials~\cite{boretti}.

\section{Conclusions}

We have performed an extensive study in how the selection of the ionic model can affect the water flow and ionic rejection by MoS$_2$ membranes. We employed two water models from the rigid TIP4P family: the traditional and well-established TIP4P/2005, and the TIP4P/$\epsilon$, recently proposed to provide the correct value of the bulk water dielectric constant. For the salt model we chose the model proposed by Joung~\cite{doi:10.1021/jp8001614}, namely NaCl/J, and the NaCl/$\epsilon$~\cite{doi:10.1021/acs.jpcb.5b12584}. The second salt model, combined with the TIP4P/$\epsilon$ water, can reproduce the dielectric constant of water and salt mixtures.

Our simulations indicate that the water and ion permeation through the nanopores is more sensitive to the ion model than to the water model employed. In fact, the screening proposed in the NaCl/$\epsilon$  leads to the ionic blockage of the nanopore with a small diameter. This mechanism was not observed previously. Also, the water around NaCl/J ion is more structured, which influences the ion entrance in the pore. 

These results indicate that distinct mechanisms can occur depending on the salt model. Not only distinct quantitative results but completely different physical behaviors. Besides that, it is well known that the next generation membrane materials for desalination technology must be very selective and fouling resistant~\cite{Werber2016}. In order to clarify this point is necessary an experimental investigation in MoS$_2$ nanopores with a diameter comparable with the ion diameter - so the ion has to be dehydrated to penetrate the pore. 

\section*{Conflicts of interest}
 There are no conflicts to declare.

\section*{Acknowledgements}
This work is financially supported by the CNPq. We thank the CENAPAD/SP, CESUP/UFRGS and SATOLEP/UFPel for the computer time. JRB thanks the Brazilian agencies CNPq and FAPERGS for the financial support.

\bibliographystyle{unsrt}  
\bibliography{references.bib}


\end{document}